\documentclass[a4paper, 12pt, openany, oneside]{article}
\usepackage[cp1251]{inputenc}
\usepackage[english]{babel}
\usepackage{ graphics,amsfonts, amsmath,bm,amssymb, array, eucal}
\usepackage[dvips]{graphicx}
\usepackage[flushleft,small]{caption2}

\makeatletter
\newcommand{\diag}{\rm \diag\, }

\renewcommand{\Re}{\mathop{\rm Re\,}}

\makeatother {

\begin{document}

\thispagestyle{empty} 
 \renewcommand{\abstractname}{\ }
\renewcommand{\refname}{\begin{center} REFERENCES\end{center}}

 \begin{center}
\bf Interaction of Electromagnetic S--Wave with a Metal
Film Located Between Two Dielectric Mediums
\end{center}\medskip
\begin{center}
  \bf  A. V. Latyshev\footnote{$avlatyshev@mail.ru$} and
  A. A. Yushkanov\footnote{$yushkanov@inbox.ru$}
\end{center}\medskip

\begin{center}
{\it Faculty of Physics and Mathematics,\\ Moscow State Regional
University, 105005,\\ Moscow, Radio str., 10--A}
\end{center}\medskip

\begin{abstract}
Generalization of the theory of interaction of  electromagnetic
$S$ -- wave  with a metal film on a case of the
film concluded between two dielectric environments is developed.

\medskip

structures
\end{abstract}

\begin{center}
{\bf Introduction}
\end{center}

Problem of interaction of an electromagnetic wave with a metal
film long time draws to itself attention \cite {F69} --
\cite {ly1}. It is connected  with theoretical interest to this
problem, and with numerous practical applications.

Nowadays there is the theory of interaction of an electromagnetic wave
 with plasma layer
in the case when electrons reflection from a film surface
has a specular character \cite {F69} --
\cite {F09}.  In these works it was  considered the freely
hanging films in air. In other words  it was considered the case, when
dielectric permeability of the environments surrounding a film is equal
to unit.

However in overwhelming majority of cases it is not so
\cite {Dressel}. As a rule in practice one deals with the films located on
some dielectric substrate. May be also
cases when the metal film is located between two
dielectric environments. Generalisation of the available theory
of interaction of electromagnetic radiation with
metal film on a such situations
will be the purpose of our work.

\begin{center}
{\bf  Problem formulation}
\end{center}

Let us consider the thin layer of metal located between two dielectric
environments. We will assume, that these environments are not magnetic.
Their dielectric permeability we will designate through $\varepsilon_1$
and $\varepsilon_2$.
Let us designate these environments by the first and the second media according to.
We consider the case, when  the first media are not absorbing.
Let's on a film from the first media the electromagnetic wave falls.
Incidence angle  we will designate $\theta$.
Let us assume, that a vector of electric field of the electromagnetic
waves is parallel to a layer surface. Such a wave is called H --
wave \cite {K} (or S -- wave \cite {F69}).

We take the Cartesian system of coordinates with the beginning of
coordinates on the surface of a layer adjoining on the first media.
We will direct axis $x$   into a layer.
And we will direct axis $y$   parallel to electric field vector
of electromagnetic wave.

Components of electric and magnetic field vectors we will
search in the form
$$
E_y(x,z,t)=E_y(x)e^{-i\omega t+ik_zz},
$$
and
$$
H_x(x,z,t)=H_x(x)e^{-i\omega t+ik_zz},\quad
H_z(x,z,t)=H_z(x)e^{-i\omega t+ik_zz}.
$$

We denote the thickness of the layer by $d$.

Let us designate by $Z^{(1)}$ an impedance on the bottom
layer surfaces at antisymmetric on electric field
configurations of external fields. It is the case 1, when
$$
E_y(0)=-E_y(d),\qquad H_z(0)=H_z(d),\qquad
\dfrac{d E_y(0)}{d x}=\frac{d E_y(d)}{d x}.
\eqno{(1)}
$$

Let us designate by $Z^{(2)}$ an impedance on the bottom
layer surfaces at symmetric on electric field
configurations of external fields. It is the case 2, when
$$
E_y(0)=E_y(d),\qquad H_z(0)=-H_z(d),\qquad
\dfrac{d E_y(0)}{d x}=-\dfrac{d E_y(d)}{d x}.
\eqno{(2)}
$$

Out of a layer it is possible to present electric fields in
the following form
$$
E_y^{(j)}(x)=\left\{\begin{array}{ll}
a_{j}h_{j}e^{ik_{2x}(x-d)}+b_{j}h_{j}e^{-ik_{2x}(x-d)},&
x>d, \\
h_{j}e^{ik_{1x}x}+p_{j}h_{j}e^{-ik_{1x}x},&
x<0, \quad j=1,2.
\end{array}
\right.
\eqno{(3)}
$$

Indexes "$1$" \, and "$2$" \, at factors $a, b, h, p $ and
field projections $E_y(x)$ correspond to the first
and to the second cases accordingly.

Thus the impedance  in both cases is defined as follows
(see, for example, \cite{S} and \cite{Landau8})
$$
Z^{(j)}=\dfrac{E_y^{(j)}(-0)}{H_z^{(j)}(-0)},\qquad j=1,2.
$$

\begin{center}
  \bf Transmittance, reflectance and absorptance
\end{center}

From Maxwell equations follows that \cite{Landau8}
$$
Z^{(j)}=\dfrac{i\omega}{c}\dfrac{ E_y^{(j)}(-0)}
{\dfrac{dE_y^{(j)}(-0)}{dx}},\qquad j=1,2.
$$

Here $c$ is the speed of light.

The symmetry of a field for the first case with use (1) and (3)
leads to the following relations
$$
-a_1-b_1=1+p_1,\qquad k_{2x}a_1-k_{2x}b_1= k_{1x}-k_{1x}p_1.
\eqno{(4)}
$$

Solving system (4), we have
$$
a_1=\dfrac{k_{1x}}{2k_{2x}}(1-p_1)-\dfrac{1+p_1}{2},
\qquad
b_1=-\dfrac{k_{1x}}{2k_{2x}}(1-p_1)-\dfrac{1+p_1}{2}.
\eqno{(5)}
$$

The  symmetry of a field for the second case with use (2) and (3)
leads to the following relations
$$
a_2+b_2=1+p_2,\qquad k_{2x}a_2-k_{2x}b_2=-k_{1x}+k_{1x}p_2.
\eqno{(6)}
$$

The solution of the system (6) has the following form
$$
a_2=-\dfrac{k_{1x}}{2k_{2x}}(1-p_2)+\dfrac{1+p_2}{2},
\qquad
b_2=\dfrac{k_{1x}}{2k_{2x}}(1-p_2)+\dfrac{1+p_2}{2}.
\eqno{(7)}
$$

Let us consider the following configuration of the field
$$
E_y(x)=b_2h_2E_y^{(1)}(x)-b_1h_1E_y^{(2)}(x).
$$

The field $E_y(x)$ has the following structure
$$
E_y(x)=\left\{\begin{array}{ll}
(a_1b_2-a_2b_1)h_1h_2e^{ik_{2x}(x-d)},&
x>d, \\
(b_2-b_1)h_{1}h_2e^{ik_{1x}x}+(p_{1}b_2-p_2b_1)h_{1}h_2e^{-ik_{1x}x},&
x<0.
\end{array}
\right.
$$

Thus, electric field corresponds to the electromagnetic
wave falling on film from negative semispace. A wave partially
passes through a film, and it is partially reflected.

Thus taking into account the equations (5) and (7) we have
$$
a_1b_2-a_2b_1=\dfrac{k_{1x}}{k_{2x}}(p_2-p_1),
$$
$$
b_2-b_1=\dfrac{k_{1x}}{k_{2x}}(1-\bar p)+1+\bar p,
$$
$$
p_1b_2-p_2b_1=\bar p+p_1p_2+\dfrac{k_{1x}}{k_{2x}}(\bar
p-p_1p_2),
$$
where
$$
\bar p=\dfrac{p_1+p_2}{2}.
$$

The quantities $p_1, p_2$ may be presented as
$$
p_{j}=\dfrac{ck_{1x}Z^{(j)}- \omega}{ck_{1x}Z^{(j)}+
\omega},\qquad j=1,2.
$$

The reflection coefficient of an electromagnetic wave is equal to
$$
R=\Big|\dfrac{\bar p(k_{1x}+k_{2x})+
2(k_{2x}-k_{1x})p_1p_2}{k_{1x}+k_{2x}+(k_{2x}-k_{1x})\bar p}\Big|^2.
$$

We may express the quantity $k_{2x}$ through $k_{1x}$ and
dielectric permeabilities $\varepsilon_1$ and $\varepsilon_2$
\cite {Landau8}
$$
k_{2x}=\dfrac{ \omega}{c}
\sqrt{\varepsilon_2-\varepsilon_1\sin^2 \theta},
\qquad k_{1x}=\dfrac{ \omega}{c}\sqrt{\varepsilon_1}\cos \theta.
$$

We can present the relations for $p_1$ and $p_2$ considering last
expression in the following form
$$
p_{j}=\dfrac{\sqrt{\varepsilon_1}\cos \theta
Z^{(j)}- 1}{\sqrt{\varepsilon_1}\cos \theta Z^{(j)}+
1},\qquad j=1,2.
$$

Average  value of  energy flux of the electromagnetic
wave $\langle{\bf S}\rangle$ is equal to \cite{Kizel}
$$
\langle{\bf S}\rangle=
\dfrac{c}{16\pi}\big\{[{\bf E}{\bf H}^*]+[{\bf E}^*{\bf
H}]\big\}.
$$

Here the asterisk designates complex conjugation.

According to Maxwell equations the magnetic field of the wave $ {\bf H} $
may be  presented in the form  \cite{Kizel}
$$
{\bf H}=\dfrac{c}{ \omega}[{\bf k}{\bf E}].
$$

According to these relations for quantity $\langle{S_x}\rangle$
we obtain the result
$$
\langle{ S_x}\rangle=\dfrac{c}{16\pi }(E_yH_z^*+E_y^*H_z)=
\dfrac{c^2}{8\pi \omega}|E_y|^2\Re(k_x).
$$

The reflection coefficient $R$ thus can be written down as
$$
R=\Bigg|\dfrac{\sqrt{\varepsilon_2-\varepsilon_1\sin^2 \theta}
(\bar p+p_1p_2)+\sqrt{\varepsilon_1}\cos \theta
(\bar p-p_1p_2)}{\sqrt{\varepsilon_2-
\varepsilon_1\sin^2 \theta}
(1+\bar p)+\sqrt{\varepsilon_1}\cos \theta(1-\bar p)}\Bigg|^2.
$$

Let us introduce the quantity
$\varepsilon_{12}=\dfrac{\varepsilon_2}{\varepsilon_1}$,
characterising the ratio of coefficients $\varepsilon_2$ and
$\varepsilon_1$. Then the relation for reflection coefficient
will be rewrited in the form
$$
R=\Bigg|\dfrac{\sqrt{\varepsilon_{12}-\sin^2 \theta}
(\bar p+p_1p_2)+\cos \theta(\bar p-p_1p_2)}{\sqrt{\varepsilon_{12}-
\sin^2 \theta}(1+\bar p)+\cos \theta(1-\bar p)}\Bigg|^2.
\eqno{(8)}
$$

We can to present the coefficient of transmission $T$ in the
form
$$
T=\dfrac{\Re(k_{2x})}{k_{1x}}\Big|\dfrac{a_1b_2-a_2b_1}{b_2-b_1}\Big|^2.
$$

The last expression may be rewrited in the form
$$
T=k_{1x}\Re(k_{2x})\Big|\dfrac{p_2-p_1}
{k_{1x}(1-\bar p)+k_{2x}(1+\bar p)}\Big|^2,
$$
or in the explicit form
$$
T=\cos \theta\Re\big(\sqrt{\varepsilon_{12}-\sin^2 \theta}\,\big)
\left|\dfrac{p_2-p_1}
{\sqrt{\varepsilon_{12}-\sin^2 \theta}(1+\bar p)+
\cos \theta(1-\bar p)}\right|^2.
\eqno{(9)}
$$

For transparent environments the quantities $\varepsilon_1$ and
$\varepsilon_2$ are real. From the obtained formula for
transmission coefficient it is clear, that in this case, when
$$
\sin^2\theta \geqslant \dfrac{\varepsilon_2}{\varepsilon_1}
$$
the transmission coefficient is equal to zero, as thus
$$
\Re\big(\sqrt{\varepsilon_{12}-\sin^2 \theta}\,\big)=0.
$$

It corresponds to full internal reflection.

We note that by $\sin^2 \theta\to\varepsilon_2/\varepsilon_1$
the transmission coefficient $T\to 0$.
The reflection coefficient  $R$ by $\theta\to \pi/2$ tends to $1$.

Now we can find the absorption coefficient  $A$ according to the formula
$$
A=1-T-R.
\eqno{(10)}
$$

Coefficients of transmission $T$ and reflection $R$ of the electromagnetic
waves by layer at $\varepsilon_1\to 1, \varepsilon_2\to 1$
transforms in earlier known expressions \cite{F69}, \cite{F66}
$$
T=\dfrac{1}{4}\big|p_{1}-p_{2}\big|^2,\qquad
R=\dfrac{1}{4}\big|p_{1}+p_{2}\big|^2.
$$

Let us consider the case of specular reflection of electrons  from a film
surface. Then for the quantities $Z^{(j)}\;(j=1,2)$
the following relations are valid \cite{F69}
$$
Z^{(j)}=\dfrac{2i\Omega}{W}
\sum\limits_{n=-\infty}^{n=\infty}
\dfrac{1}{\Omega^2\varepsilon_{tr}-Q^2},\qquad j=1,2,
$$
where
$$
W=W(d)=\dfrac{\omega_pd}{c}\cdot 10^{-7}.
$$

And the thickness of the film $d $ is measured in nanometers,
for $Z^{(1)}$ summation is conducted on odd $n$, and for
$Z^{(2)}$ on the even.

Here
$$\varepsilon_{tr}=\varepsilon_{tr}(q_1, \Omega),\qquad
\Omega=\dfrac{\omega}{\omega_p}, \qquad
\mathbf{q}_1=\dfrac{v_F}{c}\mathbf{Q},
$$
$$
\mathbf{Q}=\Big(Q_x,0,Q_z\Big),\qquad
Q_x=\dfrac{\pi n}{W(d)}, \qquad Q_z=\sqrt{\varepsilon_1}\,\Omega
\sin \theta,
$$
$\mathbf{q}_1$ is the dimensionless wave vector, $\mathbf{q}=
\dfrac{\omega_p}{v_F}\mathbf{q}_1$ is the dimensional wave vector,
$\varepsilon=\dfrac{\nu}{\omega_p}$ and $\nu$ are dimensionless and
dimensional frequencies of electron collisions accordingly,
module of vector $\mathbf{q}_1$ is equal to
$$
q_1=\dfrac{v_F}{c}\sqrt{\dfrac{\pi^2 n^2}{W^2(d)}+
\varepsilon_1\Omega^2\sin^2\theta}.
$$

Here $\omega_p$ is the plasma (Langmuir) frequency,
$\varepsilon_{tr}$ is the transverse dielectric permeability,
which may be presented in explicit form by formula
$$
\varepsilon_{tr}=
1-\dfrac{3}{4\Omega q_1^3}
\Bigg[
2(\Omega+i\varepsilon)q_1+
\Big[(\Omega+i\varepsilon)^2-
q_1^2\Big]
\ln\dfrac{\Omega+i\varepsilon-q_1}
{\Omega+i\varepsilon+q_1}\Bigg].
$$

We transform now the functions $Z^{(1)}$ and $Z^{(2)}$
$$
Z^{(1)}=\dfrac{4i\Omega}{W(d)}\sum\limits_{n=1}^{+\infty}
\dfrac{1}{\Omega^2
\varepsilon_{tr}(\Omega,\varepsilon,d,2n-1,\theta,\varepsilon_1)-
Q(\Omega,d,2n-1,\theta,\varepsilon_1)},
$$
$$
Z^{(2)}=\dfrac{2i\Omega}
{W(d)\Big[\Omega^2
\varepsilon_{tr}(\Omega,\varepsilon,d,0,\theta,\varepsilon_1)-
Q(\Omega,d,0,\theta,\varepsilon_1)}+
$$
$$
+\dfrac{4i\Omega}{W(d)}\sum\limits_{n=1}^{+\infty}
\dfrac{1}{\Omega^2
\varepsilon_{tr}(\Omega,\varepsilon,d,2n,\theta,\varepsilon_1)-
Q(\Omega,d,2n,\theta,\varepsilon_1)}.
$$

\begin{center}
  \bf Analys of results
\end{center}

We consider a fim of sodium. Then \cite{F69}
$\omega_p=6.5\times 10^{15}\sec^{-1}$, $v_F=8.52\times 10^7$ cm/sec.

Let us carry out graphic research of coefficients of transmission,
reflection and absorption,
 using  formulas (9), (8)
and (10).

Let us fix dimensionless frequency of electron collisions
for all figures $ \varepsilon=0.001$. It means,
that dimensional frequency of electron collisions  is equal:
$\nu=0.001\omega_p$.

On Figs. 1 -- 6 the normal falling of an electromagnetic wave on a
film from the first the dielectric environment is considered,
on Fig. 7 dependence of reflectance on a thickness of a film
is considered, on Figs. 8 -- 10
dependence of coefficients $T, R, A $ on an angle of
incidence of an electromagnetic wave is considered.

On Figs. 1 -- 3 the system air -  film - glass is considered
($ \varepsilon_1=1, \varepsilon_2=4$).

On Fig. 1 dependence of
transmittance coefficient on frequency
of an electromagnetic wave $T=T(\Omega,d)$ is represented. We
consider this coefficient at the various
values of a thickness of the film: $d=100$ nm (curve 1),
$d=150$ nm (curve 2) and $d=200$ nm (curve 3).

Plots on Fig. 1 show the monotonous increasing of transmittance coefficient
at frequencies less than plasma frequency: $\omega<\omega_p $, i.e.
at $ \Omega <1$. In region of the resonant
frequencies (i.e. at $ \Omega> 1$) transmittance coefficient
has oscillatory (nonmonotonic) character.
Than the thickness of the film is more, the oscillatory character is
more strongly expressed, i.e.
 local maxima and minima alternate more often.

On Fig. 2  dependence of reflectance coefficient on dimensionless
frequency of electromagnetic wave at the same values of film
thickness is represented.

Curves 1,2,3 correspond to the same values
thickness of a film, as on Fig. 2. Reflectance coefficient decreases monotonously
at $\Omega <1$ at all values of a thickness of the film. At
$\Omega>1$ we have some
local extrema. Their locations depend on the film  thickness.

On Fig. 3 dependence of absorption coefficient  on
dimensionless frequencies $\Omega$ at the same values of a film thickness
is represented. The absorptance  has a local maximum in
neighborhood of plasma frequency, i.e. at $\Omega=1$.
 The value of a local maximum depends on the film thickness.

On Figs. 4 -- 6 dependence of coefficients $T, R, A $ on
dimensionless frequency  of an electromagnetic wave
is represented. On these Figs. the film thickness is equal to $d=100$ nm.
As the first dielectric environment the air is considered.
As the second dielectric environment (substrate)
consecutive we take glass (curve 1), mica (curve 2)
and ceramics radio engineering (curve 3).

Plots on Fig. 4 show, that with  growth quantity of
dielectric permeability the value of transmittance
decrease irrespective of dimensionless oscillation frequency of
electromagnetic wave. Besides, this coefficient  monotonously
increases before the local maximum (laying in region
$1<\Omega <2$) irrespective of quantity of dielectric
permeability.

Plots on Fig. 5 show, that with growth of the dielectric
permeability the values of reflectance increase
irrespective of dimensionless  frequency of
electromagnetic wave. The reflectance has the local
minimum in the same region of $1<\Omega <2$. Besides, this
coefficient monotonously decreases before the local minimum
irrespective of quantity of dielectric permeability.

Plots on Fig. 6 show, that near to the point $\Omega=1$
values of absorptance are close to each other independent
on quantity of dielectric permeability. Change of this
coefficient has nonmonotonic character. In region of $0<\omega<1$
the absorptance has a minimum, and in region of $1<\Omega<2$
this coefficient has a maximum. At  frequencies less than $\Omega=1$
quantities of coefficient $A$ increase with decreasing of the
 values of dielectric permeability, and at $\Omega>1$ this
situation varies on the opposite.

On Fig. 7 the dependence of reflectance as
function of thickness of a film in region of $100$\, nm
$<d<200$ \, nm is represented.
As the first the dielectric environment air is considered.
As the second the dielectric environment (substrate) sequentially
we take glass (curve 1), mica (curve 2) and ceramics radio
engineering (curve 3). Plots show, that with growth of quantity of
dielectric permeability the quantity of reflectance grows independently
of the film thickness.

In last Figs. 8 -- 10 the dependence of coefficients $T, R, A $ as functions
on angle of incidence of the electromagnetic
waves is represented. Curves 1,2,3 correspond to values of a thickness
of a film
$d=100$ nm, $150$ nm, $200$ nm. Plots in these drawings show,
that values of transmittance decrease irrespective of
thickness of a film (Fig. 8). Plots on Fig. 9 show growth
of reflectance with growth of the film thickness, and a drawing on
Fig. 10  show monotonous decrease of absorptance,
and with growth of a film thickness the values of absorptance
increase also.

\begin{center}
  \bf Conclusion
\end{center}

In the present work generalisation of the theory of interaction
of electromagnetic radiation ($S$ - waves) with a metal film on a case
the film concluded between two various dielectric environments is represented.

Formulas for transmittance, reflectance and absorptance of an
electromagnetic wave are obtained.

The graphic representation  of these coefficients as a functions of
 frequency of an electromagnetic wave, angle of
incidence of the electromagnetic
waves on the film, thickness of the film and character (quantity) of the second
 dielectric environment (substrate) is fulfilled.


\begin{figure}[t]\center
\includegraphics[width=16.0cm, height=6.5cm]{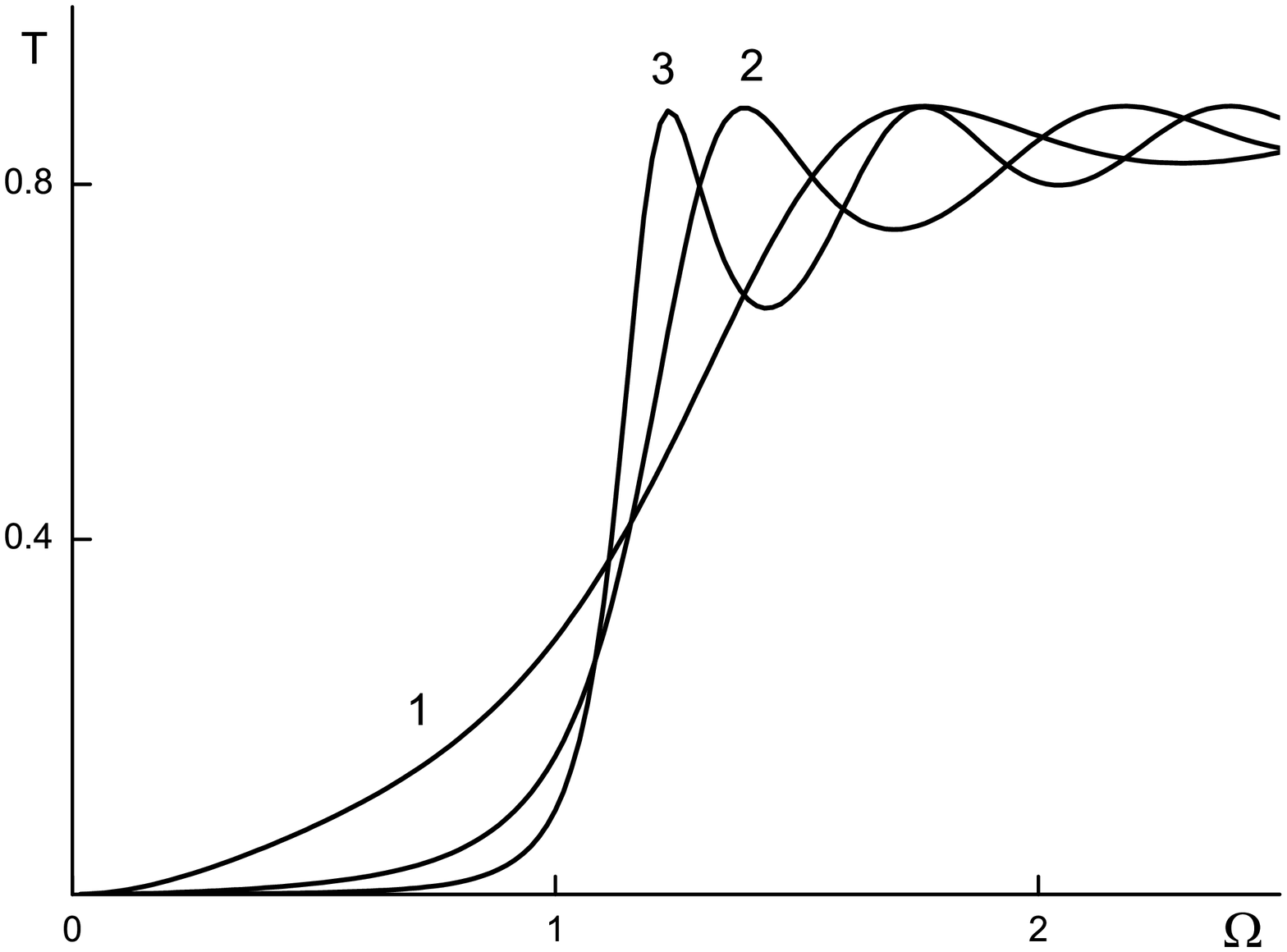}
\noindent\caption{Transmittance, air-film-glass,
curves 1,2,3 correspond to
values $d=100,150,200$ nm, $0\leqslant \Omega \leqslant 2.5$,
$\nu=0.001\omega_p$, $\theta=0^\circ$.}
\includegraphics[width=16.0cm,  height=6.5cm]{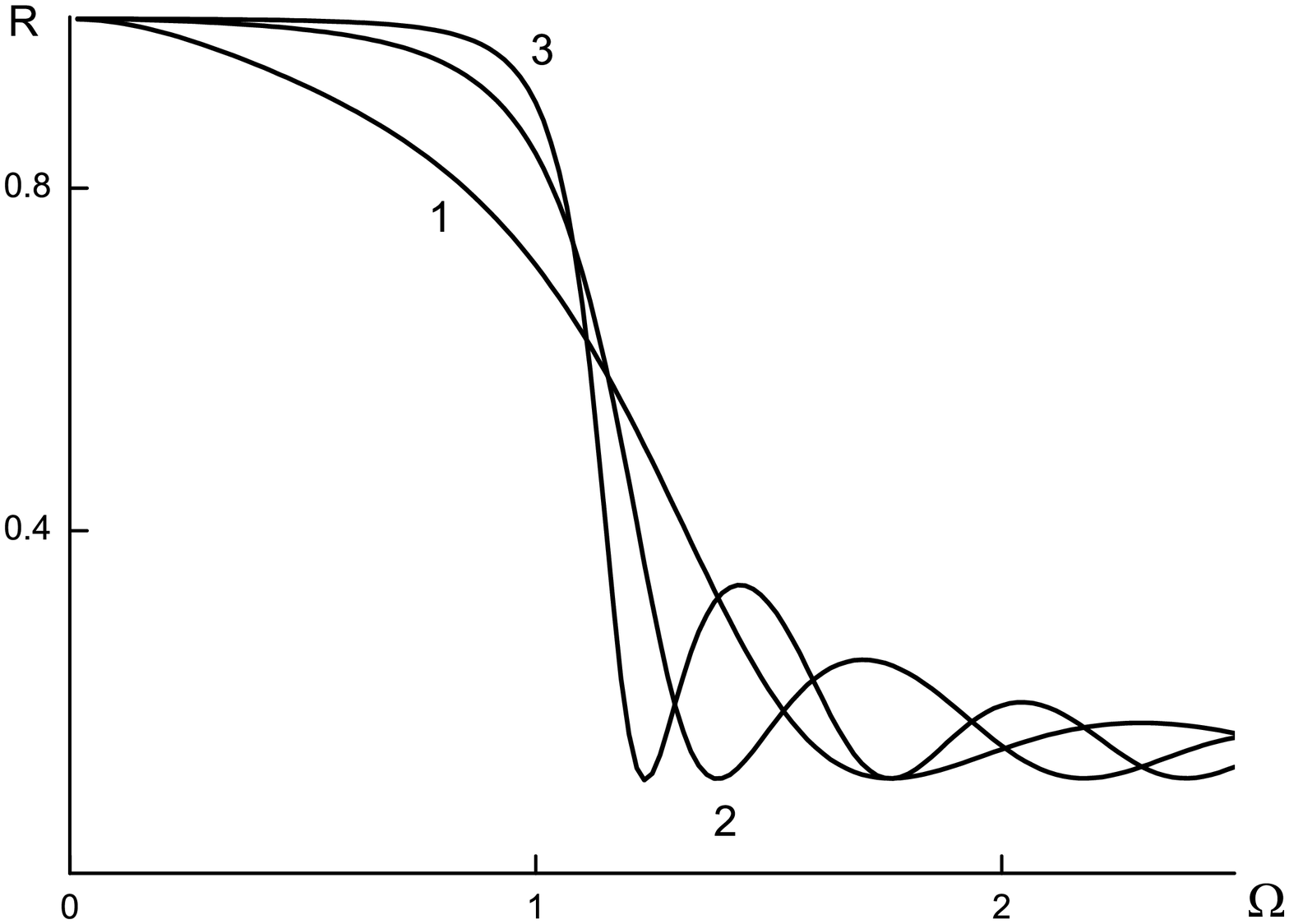}
\noindent\caption{Reflectance, air-film-glass,
curves 1,2,3 correspond to
values $d=100,150,200$ nm, $0\leqslant \Omega \leqslant 2.5$,
$\nu=0.001\omega_p$, $\theta=0^\circ$.}
\includegraphics[width=16.0cm, height=6.5cm]{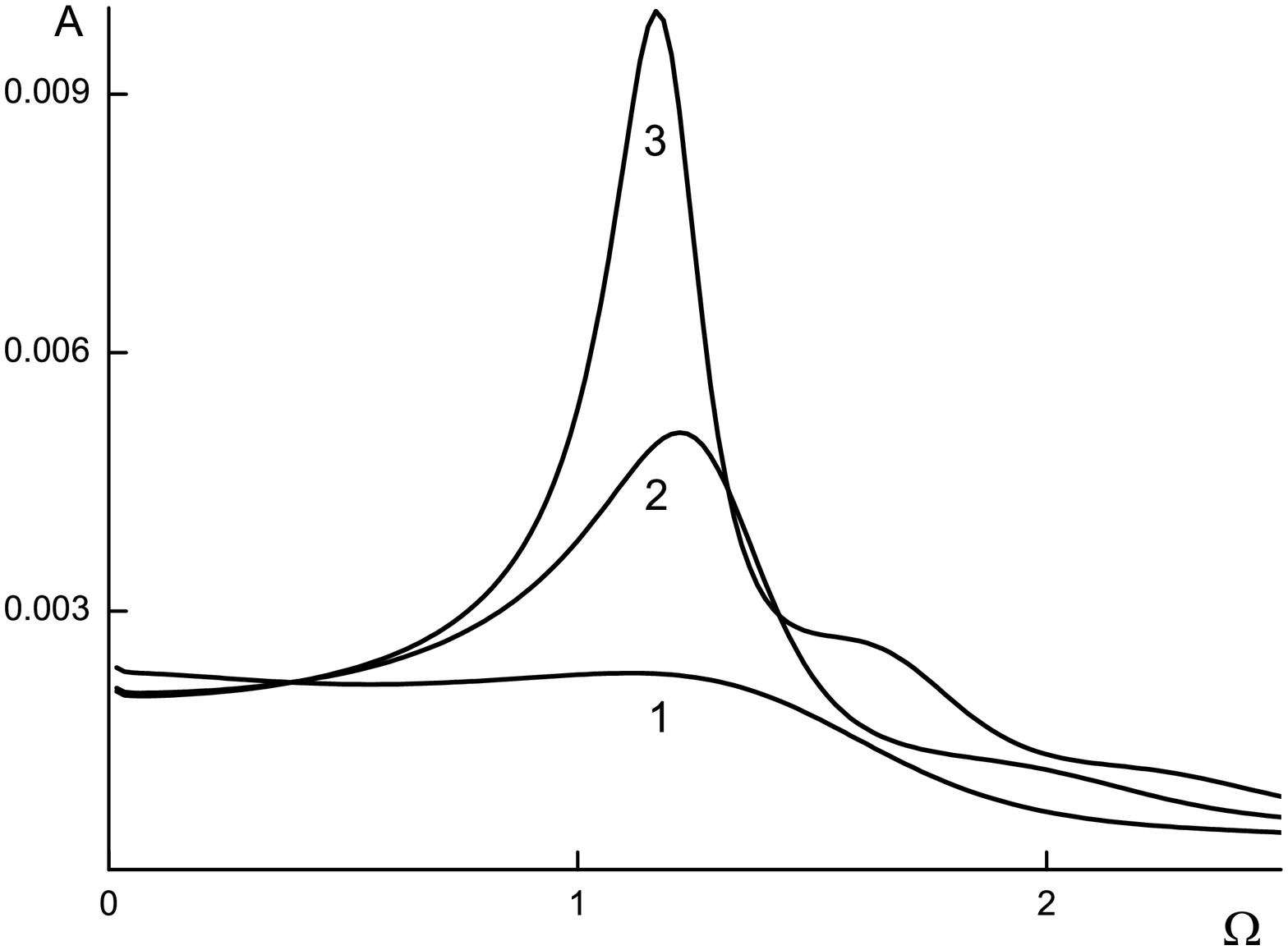}
\noindent\caption{Absortpance , air-film-glass,
curves 1,2,3 correspond to
values $d=100,150,200$ nm, $0\leqslant \Omega \leqslant 2.5$,
$\nu=0.001\omega_p$, $\theta=0^\circ$.}
\end{figure}

\begin{figure}[t]\center
\includegraphics[width=16.0cm, height=6.5cm]{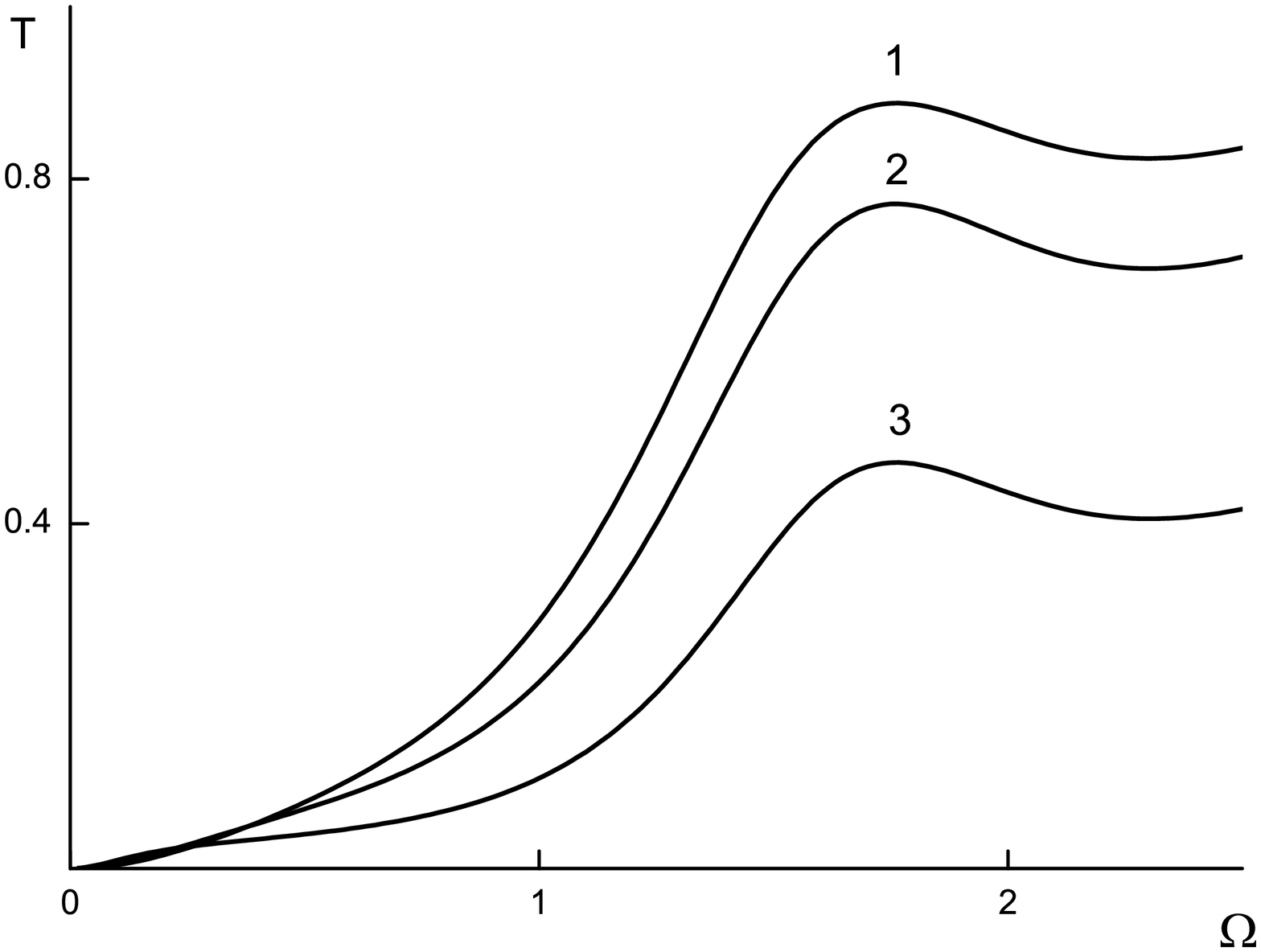}
\noindent\caption{Transmittance, $d=100$ nm, $\varepsilon_1=1$ -- air,
curves 1,2,3 correspond to
values $\varepsilon_2=4,8,40$ (glass, mica,ceramic radiotechnical),
$0\leqslant \Omega \leqslant 2.5$,
$\nu=0.001\omega_p$, $\theta=0^\circ$.}
\includegraphics[width=16.0cm,  height=6.5cm]{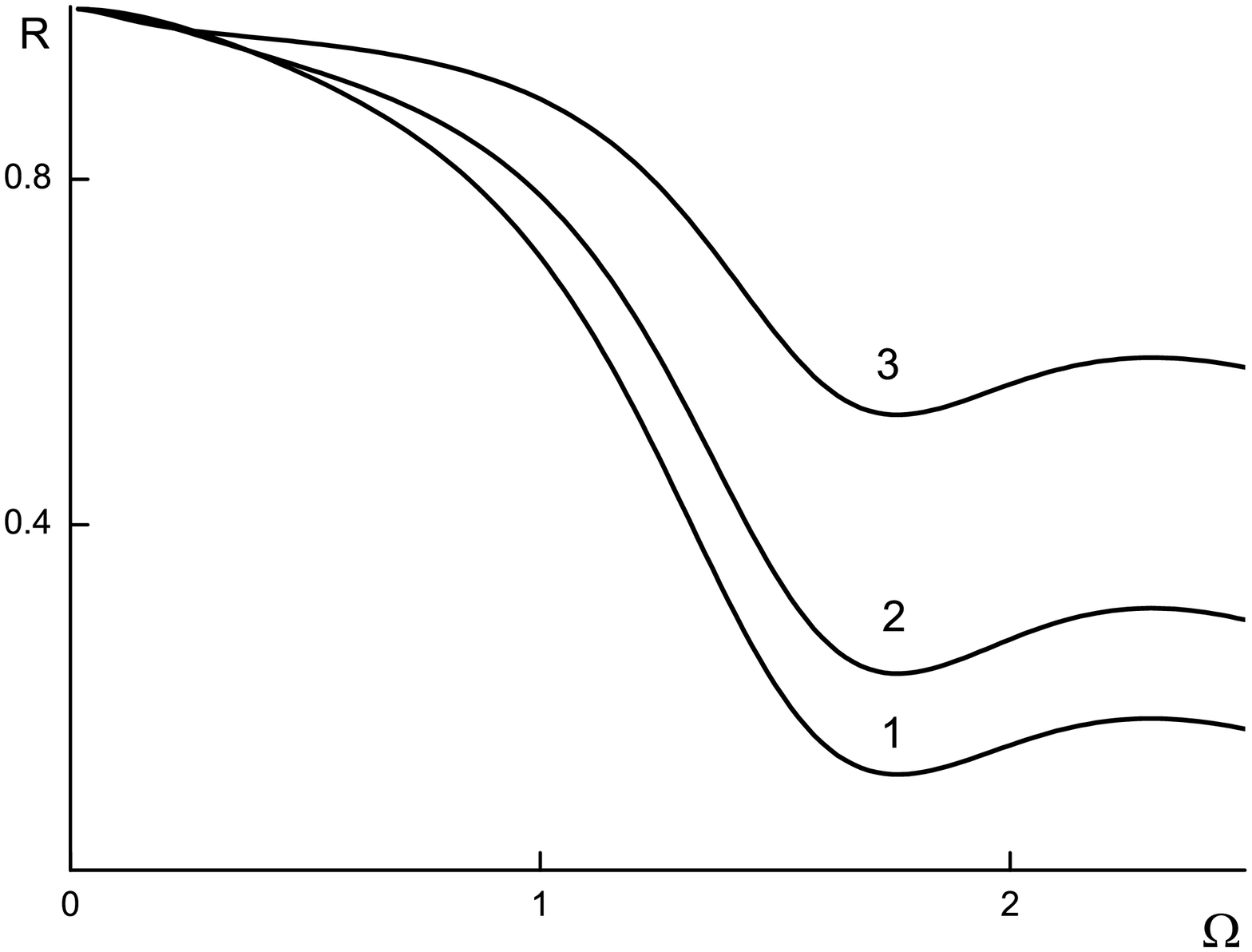}
\noindent\caption{Reflectance, $d=100$ nm,  $\varepsilon_1=1$ -- air,
curves 1,2,3 correspond to
values $\varepsilon_2=4,8,40$ (glass, mica,ceramic radiotechnical),
$0\leqslant \Omega \leqslant 2.5$,
$\nu=0.001\omega_p$, $\theta=0^\circ$.}
\includegraphics[width=16.0cm, height=6.5cm]{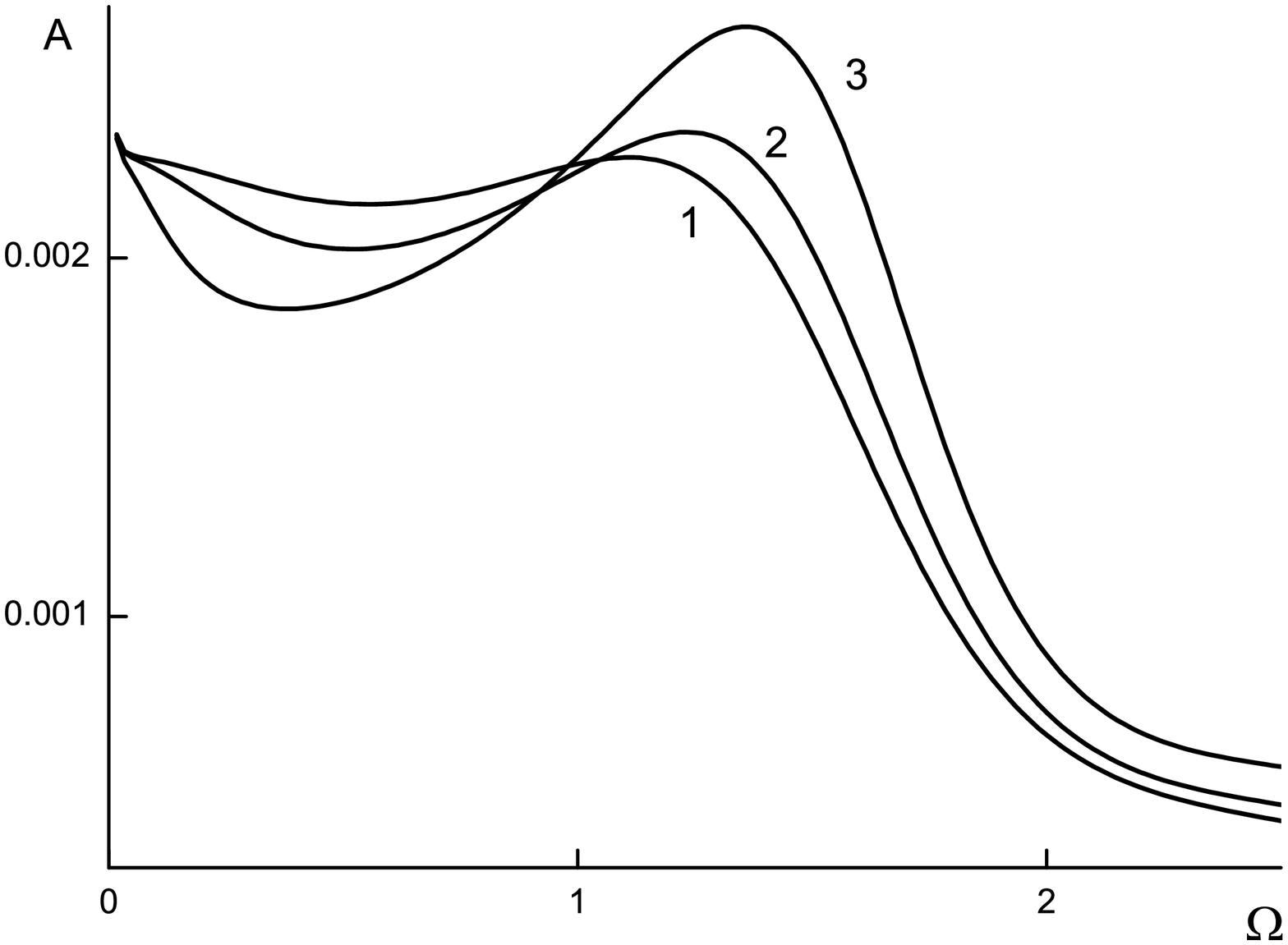}
\noindent\caption{Absortpance , $d=100$ nm,  $\varepsilon_1=1$ -- air,
curves 1,2,3 correspond to
values $\varepsilon_2=4,8,40$ (glass, mica,ceramic radiotechnical),
$0\leqslant \Omega \leqslant 2.5$,
$\nu=0.001\omega_p$, $\theta=0^\circ$.}
\end{figure}

\begin{figure}[ht]\center
\includegraphics[width=16.0cm, height=10.5cm]{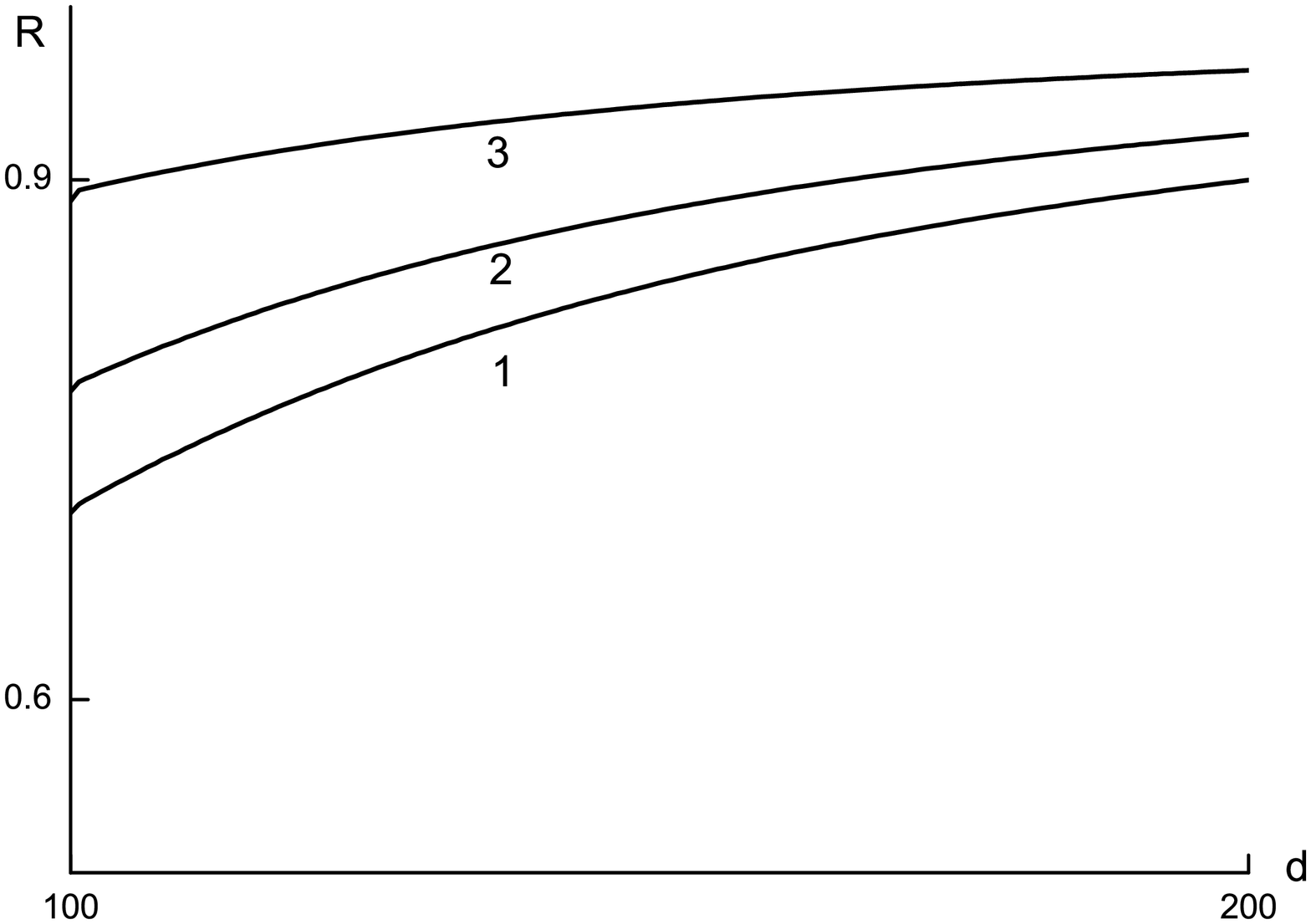}
\noindent\caption{Reflectance , $\Omega=1$,  $\varepsilon_1=1$ -- air,
curves 1,2,3 correspond to
values $\varepsilon_2=4,8,40$ (glass, mica,ceramic radiotechnical),
$100\leqslant d \leqslant 200$,
$\nu=0.001\omega_p$, $\theta=0^\circ$.}
\includegraphics[width=16.0cm, height=10.5cm]{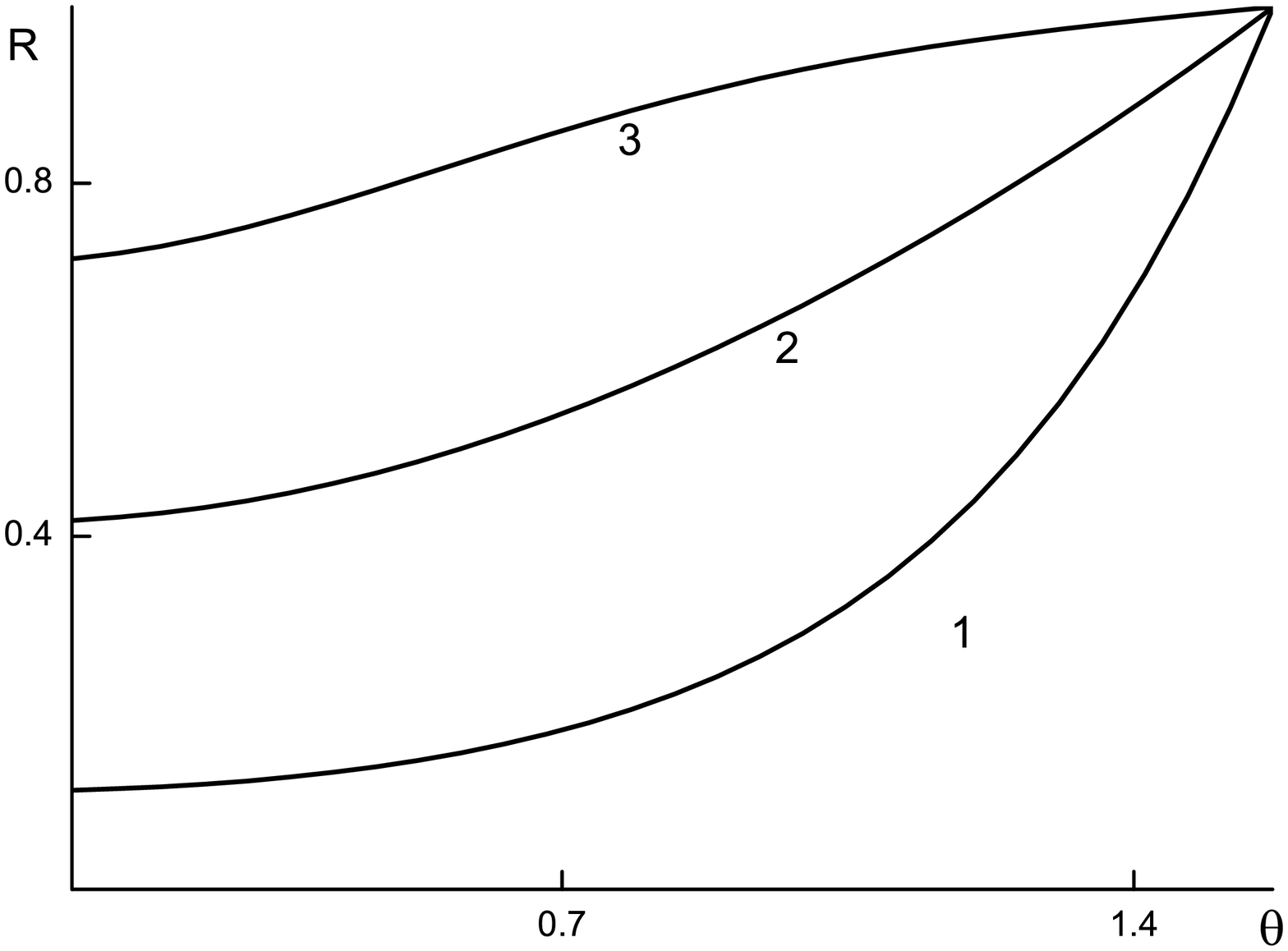}
\noindent\caption{Transmittance,  $\Omega=1$,  $\varepsilon_1=1$ -- air,
$\varepsilon_2=4$ -- glass,
curves 1,2,3 correspond to
values $d=10,50,100$,
$\nu=0.001\omega_p$, $0 \leqslant \theta \leqslant \pi/2$.}
\end{figure}
\clearpage

\begin{figure}[t]\center
\includegraphics[width=16.0cm, height=10.5cm]{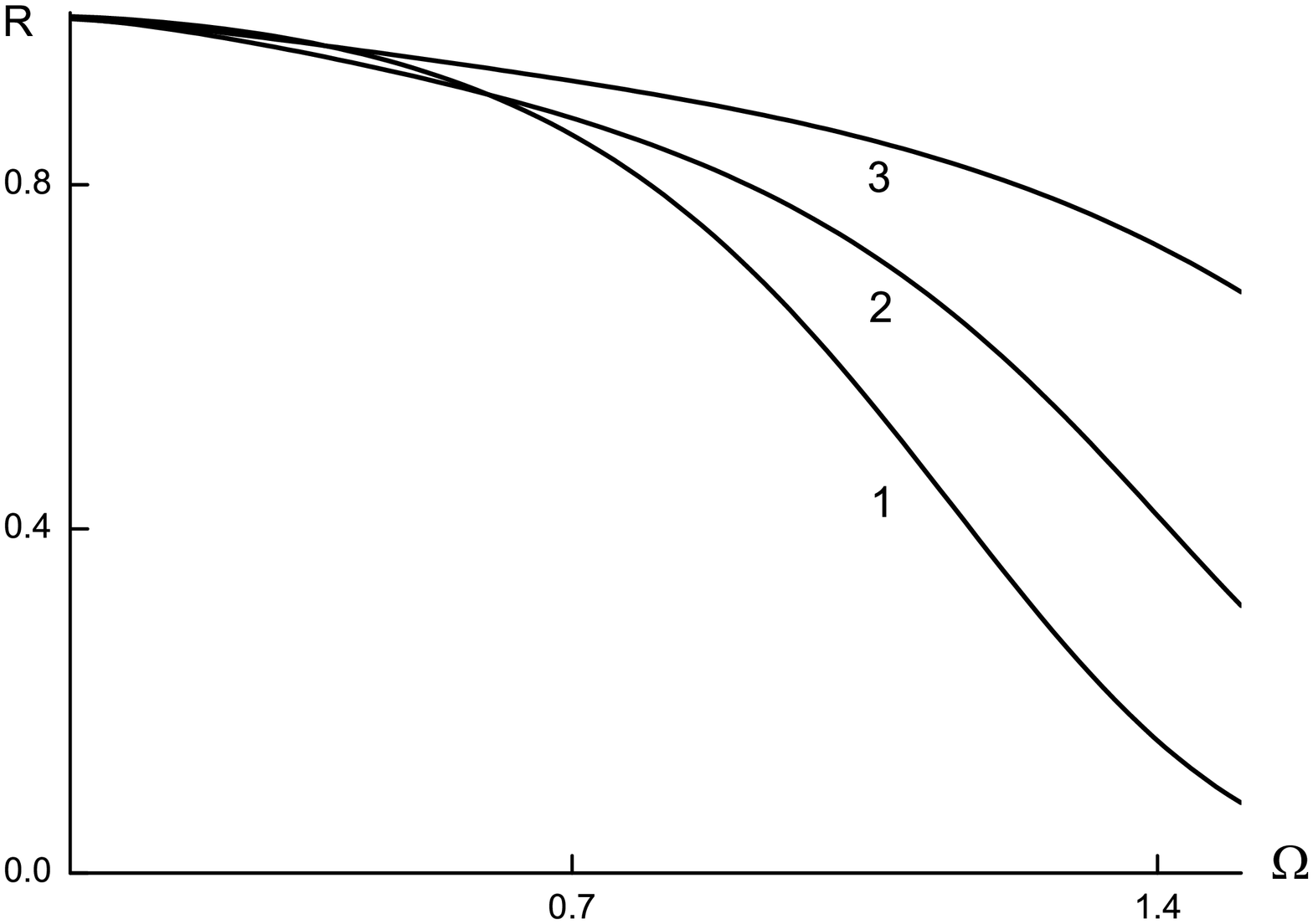}
\noindent\caption{Reflectance , $\Omega=1$,  $\varepsilon_1=1$ -- air,
$\varepsilon_2=4$ -- glass,
curves 1,2,3 correspond to
values $d=10,50,100$,
$\nu=0.001\omega_p$, $0 \leqslant \theta \leqslant \pi/2$.}
\includegraphics[width=16.0cm,  height=10.5cm]{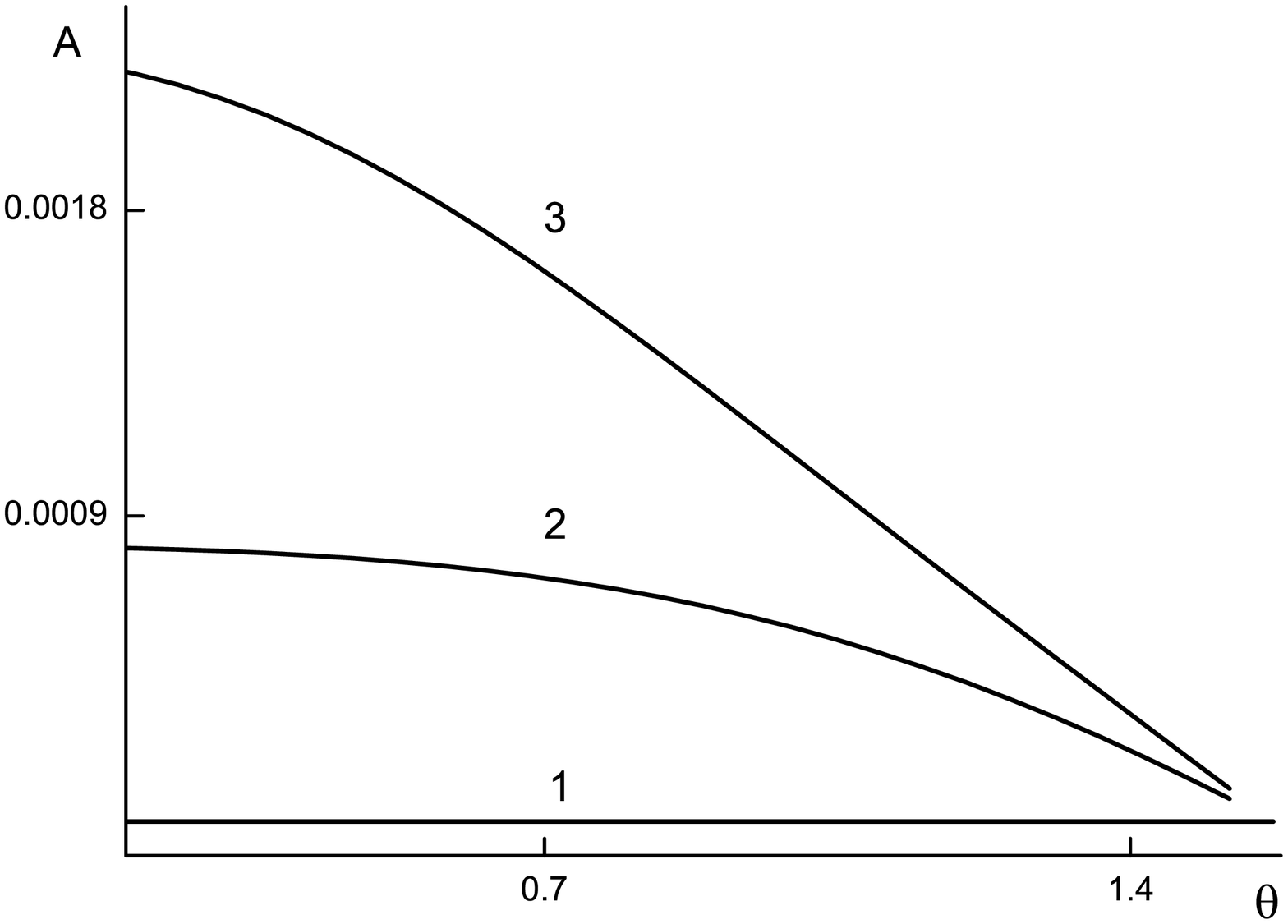}
\noindent\caption{Absorptance,  $\Omega=1$,  $\varepsilon_1=1$ -- air,
$\varepsilon_2=4$ -- glass,
curves 1,2,3 correspond to
values $d=10,50,100$,
$\nu=0.001\omega_p$, $0 \leqslant \theta \leqslant \pi/2$.}
\end{figure}


\clearpage
\begin{thebibliography}{99}\normalsize

\bibitem{F69}{\it Jones W.E., Kliewer K.L., Fuchs R.}
Nonlocal theory of the optical properties of thin metallic films
//Phys. Rev. 1969. Vol. 178. No. 3. P. 1201--1203.

\bibitem{F69-2}{\it Kliewer K.L., Fuchs R.}
Optical propertues of an electron gas: Further studies of a nonlocal
description
//Phys. Rev. 1969. Vol. 185. No. 3. P. 805--913.

\bibitem{K} {\it Kondratenko A.N.} {Penetration of waves in
plasma.} M: Atomizdat, 1979. 232 P. (in russian).

\bibitem{LY2011}
{\it Latyshev A.V. and Yushkanov A.A.}.
Interaction of the Electromagnetic S-Wave with the Thin Metal
Film
// arXiv:1010.1408 [ math-ph math.MP quant-ph], 13 pages, 5 figures.

\bibitem{F09}
{\it Paredes-Ju\'{a}rez A.,  D\'{i}as-Monge F., Makarov N.M.,
P\'{e}res-Rodr\'{i}gues F.} Nonlocal effects
in the electrodynamics of metallic slabs// JETP Lett. V. 90:9
(2010). P. 623 -- 627.


\bibitem{Dressel}
{\it Brandt T., H\"{o}vel M., Gompf B., and Dressel M.}
{ Temperature- and frequency-dependent optical properties
of ultrathin Au films.}
// Phys. Rev. B. 2008. V.78, No. 20. 205409--205415.

\bibitem{Kizel}{\it Kizel V.A.} Light Reflection, M: Science, 1973,
352 C. (in russian).

\bibitem{F66}{\it Fuchs R., Kliewer K.L., Pardee W.J.}
Optical properties of an ionic crystal slab
//Phys. Rev. 1966. Vol. 150. No.~2. P.~589--596.

\bibitem {ly1} {\it Latyshev A.V., Yushkanov A.A.}
 Exact solutions in the theory of the anomalous
 skin -- effect for a plate. Specular boundary conditions.
 Comput. Maths. Math. Phys. 1993. Vol. 33(2), p.p. 229--239.

\bibitem{Les}
{\it Latyshev A.V., Lesskis A.G., and Yushkanov A.A.}
Exact solution to the behavior of the electron plasma in a metal
layer in an alternating electric field
//Theoretical and Mathematical Physics. 1992. V. 90:2. P. 119 -– 126.

\bibitem{Landau8}
{\it Landau L.D., Lifshits E.M.}
Electrodynamics of Continuous Media,
Butterworth-Heinemann (Jan 1984). P. 460.

\bibitem{S}{\it  Sondheimer E.H.}
The mean free path of electrons in metals
// Advances in Physics. 2001. Vol. 50. No.~6. P.~499--537.

\end{thebibliography}
\end{document}